\documentclass[12pt]{iopart}

\usepackage{graphicx}
\usepackage{amssymb}

\newcommand{\structure}{{\cal S}}   
\newcommand{\base}{{b}}             
\newcommand{\ematch}{{\varepsilon_m}}       
\newcommand{\emismatch}{{\varepsilon_{mm}}} 
\newcommand{\baseinpore}{{m}}       

\begin{document}

\title{Anomalous scaling in nanopore translocation of structured heteropolymers}

\author{Malcolm McCauley${}^1$, Robert Forties${}^1$, Ulrich Gerland${}^2$ and
Ralf Bundschuh${}^3$}

\address{${}^1$Department of Physics, The Ohio State University, 191
  West Woodruff Avenue, Columbus, Ohio 43210\--1117, USA}
\address{${}^2$Arnold-Sommerfeld Center for Theoretical Physics and Center for Nanoscience (CeNS), LMU M\"unchen, Theresienstrasse 37, 80333 M\"unchen, Germany\ead{gerland@lmu.de}}
\address{${}^3$Departments of Physics and Biochemistry and
Center for RNA Biology, The Ohio State University, 191
  West Woodruff Avenue, Columbus, Ohio 43210\--1117, USA
  \ead{bundschuh@mps.ohio-state.edu}}

\date{\today}

\begin{abstract}
  Translocation through a nanopore is a new experimental technique to
  probe physical properties of biomolecules. A bulk of theoretical and
  computational work exists on the dependence of the time to
  translocate a single unstructured molecule on the length of the
  molecule. Here, we study the same problem but for RNA molecules for
  which the breaking of the secondary structure is the main barrier
  for translocation. To this end, we calculate the mean translocation
  time of single-stranded RNA through a nanopore of zero thickness and
  at zero voltage for many randomly chosen RNA sequences. We find the
  translocation time to depend on the length of the RNA molecule with
  a power law. The exponent changes as a function of temperature and
  exceeds the naively expected exponent of two for purely diffusive
  transport at all temperatures. We interpret the power law scaling
  in terms of diffusion in a one-dimensional energy landscape with a
  logarithmic barrier.
\end{abstract}


\submitto{\PB}

\bigskip


\maketitle

\section{Introduction}

Nanopore technology has opened a completely new window for probing the
properties of polymers in general and biopolymers in
particular~\cite{KasianowitzNanopores1996,MellerDNANanopore2001,NanoporeExperimentsDekkerGroup}.
In a nanopore setup two macroscopic chambers filled with a buffer
solution are separated from each other by a wall.  Embedded into this
wall is a single nanopore, i.e., a hole with a diameter in the few
nanometer range, connecting the two chambers.  When charged polymers
are added into one chamber, an electric field applied across the
nanopore can drive these polymers through the pore one by one. Drops
in the induced counter ion current due to the occlusion of the pore by
the translocating polymer allow the translocation dynamics of
individual polymers to be observed. In recent years, this technique
has been applied extensively to study
DNA~\cite{KasianowitzNanopores1996,MellerDNANanopore2001,NanoporeExperimentsDekkerGroup,MellerSequenceDependence2000,DeamerHairpinsNanopore2001,MellerHairpinUnzipping2004,SauerBudgeHairpinUnzipping2003,MatheArinsteinMeller2006,KeyserDirectForceMeasurement,MellerSolidState2008,MellerOrientationDependence2008} and
RNA~\cite{AkesonRNATranslocation1999} molecules as well as
proteins~\cite{StefureacPeptideTranslocation2006,GoodrichPeptideTranslocation}.

The emergence of this new experimental technique has also spurred a
lot of activity on the theoretical side. There has been particular
interest in understanding the nonequilibrium statistical mechanics
associated with the translocation of unstructured, linear polymers,
e.g., single-stranded DNA in which all nucleotides are the
same~\cite{SungParkTranslocation1996,DiMarzioMandell1997,MuthukumarTranslocation1999,MuthukumarPRL2001,ChenWangLuoTranslocation,SlonkinaKolomeisky2003,MilchevBinderBhattacharya2004,MellerTranslocationDynamicsReview,ChuangKantorKardarAnomalousDynamics}. The
quantities of interest are the (experimentally measurable)
distribution of translocation times, and the asymptotic behavior of
the typical translocation time as the polymers become very long.

On the simplest level of description, the translocation of a linear
polymer is hindered by an entropic barrier. An entropic barrier
emerges since the wall separating the two chambers effectively
separates the polymer into two sections: the {\it trans} section which
has already translocated and the {\it cis} section which yet has to
translocate. Each of these sections is constrained in its motion by
the wall, and the constraint is most severe when the polymer has
translocated half way through the pore. More quantitatively, if a
polymer with sequence length $N$ is divided into sections of length
$\baseinpore$ and $N-\baseinpore$, respectively, the total number of
configurations available to this polymer is reduced (compared to a
free polymer) by the power law factors $\baseinpore^{-\gamma_u}$ and
$(N-\baseinpore)^{-\gamma_u}$~\cite{EisenrieglerPolymers}. Here, the exponent $\gamma_u$ depends on
the asymptotic statistical properties of the polymer that are affected
only by the spatial dimensionality and a possible self-avoidance
interaction ($\gamma_u=1/2$ for an ideal, noninteracting chain). As a
consequence, the entropic barrier experienced by the translocating
polymer (i.e., the difference in free energy between a polymer
that just entered the pore ($\baseinpore=1$) and a polymer with
$\baseinpore$ bases on the \textit{trans} side) has the shape
\begin{equation}\label{eq_logarithmiclandscape}
F(\baseinpore) = \gamma \, k_{B}T \, \ln [(N-\baseinpore)\baseinpore/N]
\end{equation}
with $\gamma=\gamma_u\gtrapprox1/2$.  The maximum of this barrier at
$\baseinpore=N/2$ depends logarithmically on $N$, with $\gamma_u$ as a
prefactor. Modeling the translocation process as a one-dimensional
diffusion across this entropic barrier is an appropriate description,
if the translocation process is adiabatically slow, e.g. due to
friction at the pore, such that the polymer ends on each side sample
many different configurations during the time required to translocate
a macroscopic portion of the polymer.  It has been established that if
entropy reduction is the only barrier, translocation is purely
diffusive (i.e., the translocation times scale as $N^2$) in the limit
of zero voltage and ballistic (i.e., the translocation times scale as
$N$ for long polymers) at finite voltages independent of the
characteristics of the polymer model (i.e., independent of the precise
value of the exponent
$\gamma_u$)~\cite{ChuangKantorKardarAnomalousDynamics}. However, there
is still an ongoing debate what effect the actual polymer dynamics may
have on the translocation time distributions under conditions where
the adiabatic approximation breaks down, such that the polymer
dynamics is directly coupled to the translocation
dynamics~\cite{MuthukumarPRL2001,MilchevBinderBhattacharya2004,MetzlerKlafterAnomalousTranslocation2003,KantorKardar2004,MatsuyamaTranslocationKinetics2004,PanjaBarkemaTranslocationScaling2007,ChatelainKantorKardar2008}.

Here, we focus on a different, but similarly challenging theoretical
question, which arises when the translocating polymers are structured
heteropolymers. This issue has obtained some
experimental~\cite{DeamerHairpinsNanopore2001,MellerHairpinUnzipping2004,SauerBudgeHairpinUnzipping2003}
and theoretical~\cite{SauerBudgeHairpinUnzipping2003,GerlandTranslocation,BundschuhCoupledDynamics} attention but
far less than the case of unstructured molecules.
In particular, we consider a polynucleotide, an RNA or
a single-stranded DNA, consisting of a specific sequence of individual
nucleotides, i.e. A, C, G, and U for the case of RNA. For simplicity,
we will loosely use `RNA' to refer to both RNA and single-stranded DNA
in this article, as the biochemical difference between these
polynucleotides is insignificant for the questions we address. RNA
molecules have a strong propensity to form intramolecular
Watson-Crick, i.e., G--C and A--U, base pairs. The formation of such
base pairs forces the molecules to fold into sequence-dependent
structures, which are characterized by their basepairing pattern. The
naturally evolved sequences of structural RNA's, e.g. ribosomal RNA,
are biased to stably fold into particular, functional structures,
whereas the sequences of many other RNA's, e.g. most messenger RNA's,
primarily encode information, not structure. The structural features
of this latter class can be modelled via the ensemble of random RNA
sequences~\cite{HiggsRandomRNA1996,HiggsReview,BundschuhStatisticalMechanicsRNA}. Here, we characterize the translocation dynamics of
this class, focusing on the slow translocation limit. We identify
nontrivial translocation behavior, and study the physical origin of
this behavior.

Even with a random sequence, a single RNA molecule may spend most of
the time in a dominant basepairing pattern (`glassy
behavior'~\cite{HiggsRandomRNA1996}). Or else it may sample a
promiscuous array of alternative structures with different
shapes~\cite{deGennesMoltenRNA}. The transition between these two
types of behavior occurs as a function of temperature, with low
temperatures favoring glassy
behavior~\cite{BundschuhStatisticalMechanicsRNA,PagnaniParisiRNAGlass2000,BundschuhPhasesRNA}. It
is interesting to ask whether this transition is reflected also in the
translocation behavior, and if so, how?

Generally, if a folded molecule is to translocate through a very
narrow pore that allows only single strands to pass, it has to break
its base pairs in the process. This yields a coupling between the
observed translocation dynamics and the base pairing properties of the
molecule~\cite{BundschuhCoupledDynamics}.  In this system, the
separation of the polymer into a {\it cis} and a {\it trans} section
has an additional effect, namely that bases on each side of the pore
can only pair with bases on the same side of the pore thus limiting
the possible pairing partners. On average, this restriction in the
base pairing pattern again is believed to lead to a free energy
barrier that is logarithmic in the length of the polymer (see below
and~\cite{BundschuhStatisticalMechanicsRNA}). Thus, at least
superficially, the problem of a structured RNA molecule translocating
through a nanopore is mathematically similar to the problem of
homopolymer translocation, even though the physical origin of the
logarithmic barrier is completely different in nature.  However, the
problem is deeper than this analogy suggests: while the logarithmic
barrier is insignificant for the translocation of homopolymers (see
above), we will see below that for structured heteropolymers the
translocation dynamics is drastically affected. This is a consequence
of the fact that in the structured case, the prefactor $\gamma$ of the
logarithmic barrier is both bigger in magnitude (such that it exceeds
a critical threshold) and dependent on temperature.

The rest of this manuscript is organized as follows. In
Sec.~\ref{sec-methods}, we lay out our model assumptions and the
general theoretical framework used here to describe the translocation
dynamics, review the relevant aspects of the statistical physics of
RNA folding, and then link the folding and translocation
characteristics of random RNA. In
Sec.~\ref{sec_results} we first explore the translocation dynamics of
random RNA sequences numerically, and identify an anomalous scaling of
the typical translocation time with the length of the RNA. Then, we
provide some theoretical insight into the origin of this anomalous
scaling in the discussion. Sec.~\ref{sec_conclusion} summarizes our
results and provides an outlook to future work.

\begin{figure}[tb]
\begin{indented}
\item[]\includegraphics[width=8cm]{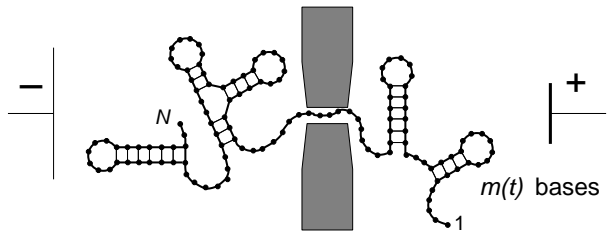}
\end{indented}
\caption{Sketch of a structured RNA molecule translocating through a
  narrow pore, which allows single but not double strands to
  pass. Translocation can be driven by an applied voltage acting on
  the negative charges of the RNA backbone. An appropriate reaction
  coordinate for the translocation process is the number of bases $m$
  that have reached the {\it trans} side. If the translocation is
  sufficiently slow, for instance due to molecular friction at the
  pore or energetic barriers caused by basepairing, $m$ becomes the
  only relevant degree of freedom. In this slow translocation limit,
  there is sufficient time for the base-pairing patterns on the {\it
    cis} and {\it trans} sides to reoptimize whenever $m$ changes.}
\label{fig1}
\end{figure}

\section{Materials and methods}
\label{sec-methods}

\subsection{Translocation dynamics: general framework}
\label{sec-translocation-1}

As illustrated in figure~\ref{fig1}, we consider a polynucleotide
translocating from the {\it cis} to the {\it trans} side of a pore in
a membrane. The pore is so narrow that only a single strand of the
polynucleotide can pass through, and hence only unpaired bases can
enter the pore.  If an external electric voltage $V$ is applied across
the pore, translocation is biased towards the positive terminal, since
RNA has a negatively charged backbone. The translocation process has a
natural ``reaction coordinate'': the number of bases $\baseinpore(t)$
that have reached the {\it trans} side at time $t$. For simplicity we
will consider an ideal pore with a negligible depth, i.e. we assume
that the remaining $N-\baseinpore$ bases are all exposed on the {\it
cis} side and none reside within the pore. In general, the
translocation process cannot be described solely by a dynamic equation
for the coordinate $m$, since the spatial and basepairing degrees of
freedom of the polymer are coupled to $\baseinpore(t)$,
see~\cite{BundschuhCoupledDynamics}. However, under conditions where
the translocation process is sufficiently slow, the translocation
dynamics becomes effectively one-dimensional, as $\baseinpore(t)$
reduces to a stochastic hopping process in an appropriate
one-dimensional free energy landscape $F(\baseinpore)$. Such a
description is appropriate if the base-pairing patterns
on the {\it cis} and {\it trans} sides have sufficient time to
reoptimize whenever $\baseinpore$ changes. Slow translocation arises
when the molecular friction at the pore is large, the voltage bias $V$
is small, and the energetic barriers due to basepairing are
significant. Throughout the present paper, we focus entirely on this
slow translocation limit.

With the above assumptions, the stochastic translocation process is described by a master equation for $P(m,t)$, the probability to find an RNA molecule with a given sequence in a state with $m$ bases on the {\it trans} side at time $t$. This master equation takes the general form 
\begin{eqnarray}
\label{master-eq}
\partial_{t} P(\baseinpore,t) &=&
k_{+}(\baseinpore\!-\!1)\,P(\baseinpore\!-\!1,t) +  \\ 
& & { } + k_{-}(\baseinpore\!+\!1)\,P(\baseinpore\!+\!1,t) + \nonumber \\
& & { } - \left[k_{+}(\baseinpore) + k_{-}(\baseinpore)\right] P(\baseinpore,t) \nonumber
\end{eqnarray}
with a set of ``hopping'' rates $k_{+}(\baseinpore)$ and $k_{-}(\baseinpore)$ that depend explicitly on the translocation coordinate $m$. Here, $k_{+}(\baseinpore)$ is the rate to translocate the base with index $\baseinpore+1$ from the {\it cis} to the {\it trans} side, whereas $k_{-}(\baseinpore)$ is the rate at which base $m$ translocates back from the {\it trans} to the {\it cis} side. 
The hopping rates also depend on the voltage bias $V$, the temperature $T$, and the nucleotide sequence of the RNA. In other words, at a given voltage bias and temperature, we need to obtain a set of $2N$ hopping rates for each RNA sequence, such that (\ref{master-eq}) describes the translocation dynamics. We then want to characterize the translocation behavior for the ensemble of random sequences. 

If the $\baseinpore$-dependence of the hopping rates is dropped, $k_{+}(\baseinpore)\equiv k_{+}$ and $k_{-}(\baseinpore)\equiv k_{-}$, (\ref{master-eq}) describes a homogeneous drift-diffusion process and becomes equivalent to the Fokker-Planck equation 
\begin{equation} 
\partial_{t} P(x,t) = D\,\partial_{x}^2 P(x,t) - v\,\partial_{x} P(x,t) 
\label{Eq_Drift_Diffusion}
\end{equation}
in the continuum limit, where $\baseinpore$ is replaced by a
continuous reaction coordinate $0<x<N$. Here, $D$ and $v$ are the
effective diffusion constant and drift velocity, respectively. As was
shown by Lubensky and Nelson~\cite{LubenskyNelsonDrivenTranslocation},
past translocation experiments with unstructured single-stranded
polynucleotides are quantitatively consistent
with~(\ref{Eq_Drift_Diffusion}): The experimental distribution of
translocation times $p(\tau)$ is well described by the corresponding
distribution from (\ref{Eq_Drift_Diffusion}), which is determined by
the probability current into the absorbing boundary at $x\!=\!N$.
 
\begin{figure}
\begin{indented}
\item[]\includegraphics[width=0.8\columnwidth]{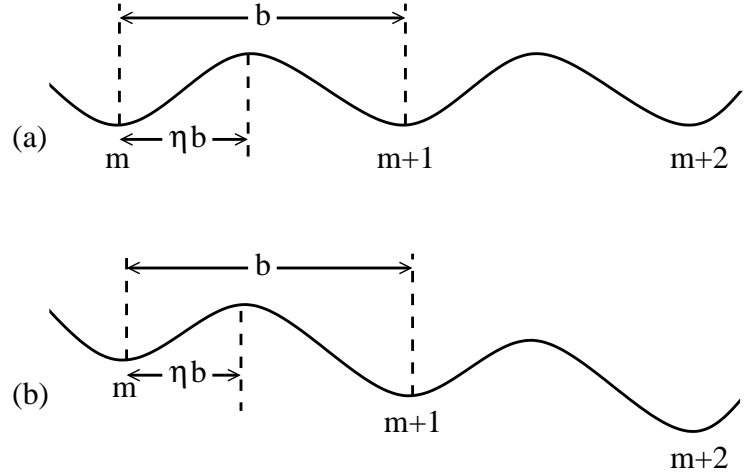}
\end{indented}
\caption{Illustration of the voltage-dependence of the translocation
  rates in (4) of the main text. Even in the absence of secondary
  structure, the translocation of a single base is envisaged as a
  barrier crossing process. The coarse-grained, discrete reaction
  coordinate $m$ (the number of translocated bases) then corresponds
  to the minima of a continuous microscopic free energy landscape. The
  distance of the minima reflects the base-to-base distance $b$ of the
  RNA. The position of the transition state, at a fractional distance
  $\eta$ from the minimum to the left, is an unknown microscopic
  parameter which determines how the biasing effect of the applied
  voltage is split between the forward and reverse translocation
  rates: When the unbiased landscape of (a) is tilted by the applied
  voltage as shown in (b), the reduction in the free energy barrier
  for forward translocation is proportional to $\eta$, while the
  increase in the barrier for reverse translocation is proportional to
  $(1-\eta)$.\label{fig_etasketch}}
\end{figure}

For structured RNA's, we express the hopping rates of (\ref{master-eq}) more explicitly in the form 
\begin{eqnarray}
\label{hopping-rates}
k_{+}(\baseinpore) &=& k_{0} \cdot w_{cis}(\baseinpore) \cdot \exp\left(\eta\,\frac{q_{\rm eff}V}{k_{B}T}\right)  \\
k_{-}(\baseinpore) &=& k_{0} \cdot w_{trans}(\baseinpore) \cdot \exp\left((\eta-1)\,\frac{q_{\rm eff}V}{k_{B}T}\right) \nonumber \;.
\end{eqnarray}
Here $k_{0}$ denotes the basic ``attempt'' rate for the translocation
of a single unpaired base, while $w_{cis}(\baseinpore)$ and
$w_{trans}(\baseinpore)$ denote the probability that the base
attempting to translocate is indeed not paired. The exponential
(Arrhenius) factors account for the voltage bias $V$ across the pore,
which acts on the effective charge $q_{\rm eff}$ of a
nucleotide~\cite{KeyserDirectForceMeasurement,ShklovskiiEffectiveCharge}
(note that the applied voltage drops primarily directly across the
pore, while the nucleotides do not experience a significant
electrostatic force on either side).  The dimensionless factor $\eta$
is a measure for the position of the microscopic transition state that
limits the rate for the crossing of a single nucleotide. More
precisely, $\eta$ is the relative distance of this transition state
from the entrance of the pore; for a symmetric pore, $\eta=1/2$ (see
figure~\ref{fig_etasketch}).

For a fixed but arbitrary set of hopping rates $k_{+}(\baseinpore)$,
$k_{-}(\baseinpore)$, the (thermal) average of the translocation time
can be calculated analytically using the mean first passage time
formalism~\cite{GardinerStochasticMethods}. One obtains
\begin{equation}
  \label{MFPT}
  \langle\tau(\baseinpore_{0})\rangle = \sum_{m=\baseinpore_{0}}^{N-1}\sum_{l=0}^{m} \frac{1}{k_{+}(l)} \prod_{j=l+1}^{m} \frac{k_{-}(j)}{k_{+}(j)} \;.
\end{equation}
This equation assumes that at time $t=0$, the translocation process
has already proceeded to the translocation coordinate
$\baseinpore(0)=\baseinpore_{0}$. While the entire translocation
process consists of an entrance stage (with possible failed attempts),
followed by a passage stage, our focus here is only on the
latter. More precisely, we are interested in the detailed passage
dynamics of the successful translocation events. Equation~(\ref{MFPT})
assumes reflecting boundary conditions at $\baseinpore=0$, i.e. the
molecule is only allowed to exit the pore on the {\it trans} side, as
in previous theoretical studies~\cite{MuthukumarPRL2001,ChuangKantorKardarAnomalousDynamics,MetzlerKlafterAnomalousTranslocation2003,KantorKardar2004}. Experimentally, this
corresponds to a situation where, e.g., a protein or a small bead is
attached to the {\it trans} end of the molecule, preventing exit to
the {\it cis} side. In particular at low driving voltages, such a
``road block'' will be experimentally required, since otherwise it
would not be possible to separate failed translocation attempts from
full translocation events to the {\it trans} side. At larger driving
voltages, the boundary condition at $\baseinpore=0$ is expected to be
less relevant, since molecules are then unlikely to exit the pore on
the {\it cis} side once they are inserted into the pore. At the other
end, $\baseinpore=N$, (\ref{MFPT}) assumes an absorbing boundary,
i.e., the translocation time $\tau$ is defined as the time when the
state $\baseinpore=N$ is first reached.

To determine the hopping rates (\ref{hopping-rates}) for an RNA molecule with a given sequence, we first need to calculate the probabilities $w_{cis}(\baseinpore)$ and $w_{trans}(\baseinpore)$. To this end, we review in the following section the physics of RNA folding and the characteristics of random RNA sequences, before we return to link these characteristics to the translocation dynamics in Sec.~\ref{sec-translocation-2}.

\subsection{Folding of random RNA sequences}
\label{sec-random-RNA}

In this section, we will review the aspects of the statistical physics
of structures of random RNA molecules that are relevant for our
study. We will follow the bulk of the previous literature and
exclusively focus on RNA secondary
structures~\cite{HiggsRandomRNA1996,HiggsReview,PagnaniParisiRNAGlass2000,BundschuhPhasesRNA,MarinariRNAZeroTemperature2002,KrzakalaMezardMuller2002,LeihanRNAGlass2006}. An
RNA secondary structure is the collection of all base pairs formed by
a molecule. Formally, it can be described as a set
$\structure=\{(i_1,j_1),\ldots,(i_n,j_n)\}$ of all pairs of indices
$(i_k,j_k)$ (with $i_k<j_k$) of bases that are paired. A pairing
configuration is only considered to be a valid secondary structure if
it fulfils two conditions: (i) Each base is paired with at most one
other base. (ii) If $(i,j)$ is a base pair and $(k,l)$ is a base pair
with $i<k$, they have to be either nested, i.e., fulfil $i<k<l<j$, or
independent, i.e., fulfil $i<j<k<l$. Forbidden base pairing
configurations with $i<k<j<l$ are called pseudo-knots. Restricting the
allowable secondary structures to only those that contain neither base
triplets nor pseudo-knots is an approximation since both structural
elements do occur in actual structures. However, the approximation is
reasonable since base triplets and pseudo-knots are believed to be
rare in natural structures and can be effectively suppressed by
performing experiments in the absence of multi-valent ions, such as
Mg$^{2+}$~\cite{TinocoHowRNAFolds}.

The energy $E[\structure]$ of a given structure $\structure$ depends
on the sequence $\base_1\ldots\base_N$ of the RNA molecule. For
quantitative analyzes such as the prediction of the actual secondary
structure of an RNA
molecule~\cite{HofackerVienna1994,MfoldServer2003}, very detailed
energy models with hundreds of parameters have been
developed~\cite{MathewsRNAParameters1999}. Since we are interested in
more generic questions such as the scaling behavior of translocation
times, we will use a strongly simplified energy model that focuses on
the base pairing alone. More specifically, we will assign an energy
solely derived from the base pairs formed in the structure, i.e.,
\begin{equation}
E[\structure]=\sum_{(i,j)\in \structure}\varepsilon_{i,j}
\end{equation}
where $\varepsilon_{i,j}$ is the energy for the formation of a base
pair between base $b_i$ and
$b_j$. Following~\cite{BundschuhStatisticalMechanicsRNA,BundschuhPhasesRNA}
we will even ignore the differences between the stability of different
Watson-Crick base pairs and use the simplest possible model
\begin{equation}\label{eq_matchmismatchenergies}
\varepsilon_{i,j}=\left\{\begin{array}{ll}-\ematch&\mbox{$b_i$ and
      $b_j$ are a Watson-Crick
      pair}\\\emismatch&\mbox{otherwise}\end{array}
\right.
\end{equation}
where the match and mismatch energies $\ematch$ and $\emismatch$ are positive constants. Such a simplified energy model clearly is not suitable for the quantitative prediction of the behavior of an individual RNA molecule. However, the universal properties of the RNA folding problem, such as the thermodynamic phases, the topology of the phase diagram, and the critical exponents characterizing these phases in the thermodynamic limit are expected to be correctly captured.

For this as well as other more complicated energy models, the
partition function of an RNA molecule of a given sequence
$\base_1\ldots\base_N$ can be calculated exactly in polynomial
time~\cite{McCaskillRNAPartitionFunction}. This can be done by
introducing as an auxiliary quantity the partition function $Z_{i,j}$
for the substrand $\base_i\ldots\base_j$ of the original molecule. The
$j$th base can either be unpaired or paired with the $k$th base, where
$k$ takes all of the possibilities from $i$ to $j-1$.  If the $j$th
base is unpaired, the allowable structures are exactly the allowable
structures for the substrand $\base_i\ldots\base_{j-1}$. If the $j$th
base is paired with the $k$th base, the exclusion of pseudo-knots
implies that in the presence of the $(k,j)$ base pair, any structure
is possible on the substrand $\base_i\ldots\base_{k-1}$ and on the
substrand $\base_{k+1}\ldots\base_{j-1}$ but base pairs between these
two substrands are forbidden. That yields the recursion equation
\begin{equation}
\label{eq_rnarecursion}
Z_{i,j}=Z_{i,j-1}+\sum_{k=i}^{j-1}Z_{i,k-1}e^{-\beta\varepsilon_{k,j}}Z_{k+1,j-1}
\end{equation}
where $\beta=(k_BT)^{-1}$. Since the substrands referred to on the
right hand side of this equation are shorter than the substrands
referred to on the left hand side, this recursion equation can be used
to start from the trivial single and two base substrands and calculate
the partition functions for the increasingly larger substrands. The
partition function $Z_{1,N}$ is then the partition function of the
whole molecule. Since in this process $O(N^2)$ of the $Z_{i,j}$ have
to be calculated with each calculation requiring one summation over
the index $k$, the total computational complexity for this calculation
is $O(N^3)$.

Through various numerical and analytical arguments it has been
established that RNA secondary structures undergo a glass transition
between a high temperature molten and a low temperature glassy
phase~\cite{HiggsRandomRNA1996,BundschuhStatisticalMechanicsRNA,PagnaniParisiRNAGlass2000,BundschuhPhasesRNA,KrzakalaMezardMuller2002,HartmannGlassComment2001,PagnaniReply2001}.  In
the (high temperature) molten phase the energetic differences between
different structures become irrelevant and configurational entropy is
the main contributor to the free energy of the structural
ensemble~\cite{deGennesMoltenRNA} (it is to be noted that our
simplified model of RNA secondary structures does not show a
denaturation transition where base pairing itself becomes unfavorable
and the molecule becomes completely unstructured. Thus, ``high
temperature'' in terms of real RNA molecules refers to temperatures
still below the denaturation temperature, but close enough so that the
energetic differences between different base pairs are smeared
out). In the glassy phase, one or a few structures (determined by the
specific sequence of the molecule) become dominant in the thermal
ensemble --- the molecule ``freezes'' into those structures.

The molten (high temperature) phase of RNA secondary structures is
completely understood analytically~\cite{deGennesMoltenRNA}. Since in
the molten phase by definition the base pairing energetics do not play
a role any more, the behavior of the molten phase can be determined by
setting all base pairing energies equal, i.e., by choosing
$\varepsilon_{i,j}=-\varepsilon_0$ with some positive
$\varepsilon_0$. Under this choice the partition functions $Z_{i,j}$
no longer depend on the nucleotide sequence and thus become
translationally invariant, i.e., $Z_{i,j}\equiv Z(j-i+1)$. The
recursion equation~(\ref{eq_rnarecursion}) then simplifies
to
\begin{equation}
Z(N+1)=Z(N)+q\sum_{k=1}^{N}Z(k-1)Z(N-k)
\end{equation}
where $q\equiv\exp(\beta\varepsilon_0)$ is the Boltzmann factor
associated with a base pair. This recursion equation can be solved in
the limit of large $N$ and yields
\begin{equation}
\label{Z-molten}
Z(N)\approx A N^{-\gamma_m} z_0^N
\end{equation}
where $A$ and $z_0$ depend on the Boltzmann factor $q$. The exponent $\gamma_m=3/2$, however, is universal and is characteristic of the molten phase.

\subsection{Translocation of random RNA sequences}
\label{sec-translocation-2}

In the context of polymer translocation, it is necessary to determine
what effect the pore has on the possible secondary structures of the
molecule. If direct interactions with the pore are ignored, the only
effect of the pore is that it divides the molecule into two segments,
namely the \textit{trans} part with $\baseinpore$ bases and the
\textit{cis} part with $N-\baseinpore$ bases. Each part of the
molecule can still form RNA secondary structures, but base pairs
between a base on the \textit{trans} side and a base on the
\textit{cis} side become impossible. This constraint results in a free
energy cost. In the entropically dominated molten phase a reduction in
the number of possibilities for base pairing will decrease the
entropy; in the energetically dominated glassy phase, a reduction in
the number of possibilities to find well matching substrands will
increase the energy. In both cases, the free energy cost provides a barrier to the translocation process, and we refer to the cost as the pinch free energy $F(\baseinpore)$. The pinch free energy depends explicitly on our reaction coordinate $\baseinpore$ and hence constitutes a free energy landscape for the translocation process\footnote{To keep the notation concise, we suppress the dependence of the pinch free energy $F(\baseinpore)$ on the total sequence length $N$.}. 

With the help of the partition function $Z_{i,j}$ introduced in the previous section, the pinch free energy can be easily calculated: 
The partition function for the RNA molecule at position $\baseinpore$ in the pore has the product form $Z_{1,m}Z_{m+1,N}$ (the structures on the \textit{cis} and \textit{trans} sides are uncorrelated), whereas the partition function of the unconstrained RNA in solution is $Z_{1,N}$. The free energy difference between these states is the pinch free energy, 
\begin{equation}
\label{pinchFE}
  F(\baseinpore) = -k_{B}T \left[\ln \left(Z_{1,m}Z_{m+1,N}\right)-\ln Z_{1,N}\right] \;.
\end{equation}
Using the definition of the partition function, we can also establish the explicit link of the pinch free energy landscape to the translocation dynamics model of Sec.~\ref{sec-translocation-1}. To this end, we need to determine the probabilities $w_{cis}(\baseinpore)$ and $w_{trans}(\baseinpore)$ in (\ref{hopping-rates}). Since $Z_{i,j}$ represents the total statistical weight of all permitted basepairing patterns for the RNA substrand from base $i$ to base $j$, the probability $w_{cis}(\baseinpore)$ for the base immediately in front of the pore on the {\it cis} side to be unpaired is given by  
\begin{equation}
\label{wcis}
w_{cis}(\baseinpore) = \frac{Z_{\baseinpore+2,N}}{Z_{\baseinpore+1,N}} \;.
\end{equation}
Similarly, the probability for the base immediately in front of the pore on the {\it trans} side to be unpaired is given by  
\begin{equation}
\label{wtrans}
w_{trans}(\baseinpore) = \frac{Z_{1,\baseinpore-1}}{Z_{1,\baseinpore}} \;.
\end{equation}
Together, (\ref{master-eq}), (\ref{hopping-rates}), (\ref{eq_rnarecursion}), (\ref{wcis}), and (\ref{wtrans}) fully specify the translocation dynamics of structured RNA molecules within our model. The general form (\ref{MFPT}) for the average translocation time then simplifies to~\cite{BundschuhCoupledDynamics} 
\begin{eqnarray}
  \label{EQmeantau}
  k_{0} \langle\tau\rangle &=& e^{-\eta\,\frac{q_{\rm eff}V}{k_{B}T}}
  \sum_{m=\baseinpore_{0}}^{N-1} \sum_{\ell=0}^{m} e^{-(m-\ell)\frac{q_{\rm eff}V}{k_{B}T}} 
  \frac{Z_{1,\ell}Z_{\ell+1,N}}{Z_{1,m}Z_{m+1,N}} \nonumber \\ 
  &=& e^{-\eta\,\frac{q_{\rm eff}V}{k_{B}T}}
  \sum_{m=\baseinpore_{0}}^{N-1} \sum_{\ell=0}^{m} 
  e^{\frac{F(m)-F(\ell)-(m-\ell) q_{\rm eff}V}{k_{B}T}}
\end{eqnarray}
using the free energy $F(\baseinpore)$ as defined in (\ref{pinchFE}).
It is now evident that the translocation dynamics of
Sec.~\ref{sec-translocation-1} corresponds to a random walk in the pinch free energy landscape which is tilted by the applied voltage. 

Equation (\ref{pinchFE}) can be used to compute the free energy landscape for a specific RNA sequence. To characterize the typical translocation behavior of structured RNA molecules, we need to generate such landscapes for a large sample from the ensemble of random sequences. We will take this numerical approach in section \ref{sec_numerics}. However, using (\ref{Z-molten}), we can analytically determine the typical form $F_{\mathrm{molten}}(\baseinpore)$ of the landscape in the molten phase,  
\begin{eqnarray}
\label{Fmolten}
F_{\mathrm{molten}}(\baseinpore)&=&
-k_BT\ln\frac{Z(\baseinpore)Z(N-\baseinpore)}{Z(N)} \nonumber \\
&\approx&-k_BT\ln\frac{\baseinpore^{-\gamma_m}(N-\baseinpore)^{-\gamma_m}}{N^{-\gamma_m}} \nonumber \\
&=&\gamma_m k_B T\ln[\baseinpore(N-\baseinpore)/N] \;.
\end{eqnarray}
This is formally the same logarithmic free energy landscape as for the translocation of unstructured polymers, (\ref{eq_logarithmiclandscape}). However, its physical origin is completely different (namely, the structural entropy of base pairing configurations rather than the positional entropy of the backbone), and its prefactor $\gamma_m=3/2$ is larger, which will be important below. 

It is interesting to note that the logarithmic behavior of
(\ref{Fmolten}) and the value of the prefactor can be physically
understood by realizing that the ensemble of secondary structures in
the molten phase corresponds to the ensemble of (rooted) branched
polymers: The number of possible configurations of a rooted branched
polymer of molecular weight $\baseinpore$ is known to scale like
$\baseinpore^{-3/2}$~\cite{LubenskyObukhovBranchedPolymers1981} (in
addition to the non universal extensive factor) and thus the pinch
free energy landscape of a translocating RNA molecule in the molten
phase is the same as the landscape generated by cutting a branched
polymer of molecular weight $N$ into two rooted branched polymers of
molecular weights $\baseinpore$ and $N-\baseinpore$,
respectively~\cite{BundschuhStatisticalMechanicsRNA}.

\begin{figure}
\begin{indented}
\item[]\includegraphics[width=0.8\columnwidth]{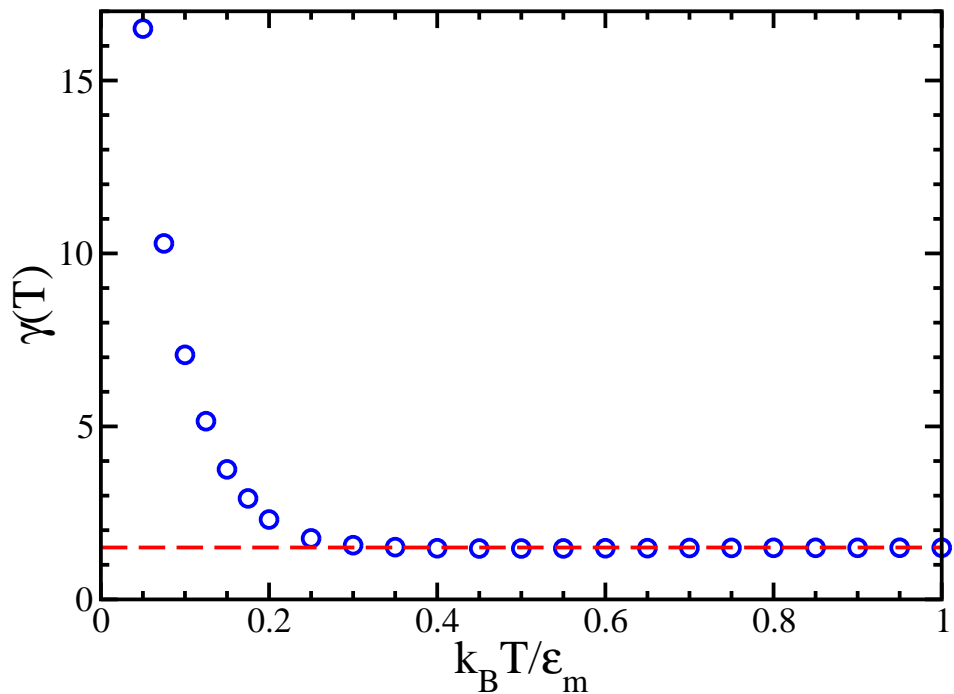}
\end{indented}
\caption{Numerically determined prefactor \protect$\gamma(T)$ of the
  logarithmic free energy landscape
  (\ref{eq_logarithmiclandscape}) as a function of temperature
  (most of the data from~\cite{BundschuhStatisticalMechanicsRNA}). The
  statistical error of the data is on the order of the symbol size.
  It can be seen that the prefactor is constant $\frac{3}{2}$ in the high
  temperature (molten) phase. In the low temperature (glassy) phase
  the prefactor becomes temperature dependent and diverges. The
  prefactors were determined by generating many random RNA sequences
  with equal probability of the four bases of lengths $N=160$, $320$,
  $640$, and $1280$, calculating the restricted partition functions
  for the energy model (\ref{eq_matchmismatchenergies}) with
  $\ematch=\emismatch$ via (\protect\ref{eq_rnarecursion}) and
  extracting the pinch free energies at $\baseinpore=N/2$ via
  (\ref{pinchFE}). The prefactor $a(T)$ of the logarithmic law for
  the pinch free energy was determined by fitting such a logarithmic
  law to the numerical data and the corresponding prefactor of the
  logarithmic free energy landscape $\gamma(T)$ was calculated from
  (\protect\ref{eq_gammafroma}).
  \label{fig_prefactorvstemperature}}
\end{figure}

In the glassy phase the situation is much less clear, since there are
no analytical calculations of the typical pinch free energy for the
ensemble of random RNA sequences. Furthermore, different numerical
studies~\cite{BundschuhStatisticalMechanicsRNA,KrzakalaMezardMuller2002,LeihanRNAGlass2006},
which examined the maximal pinch (at $\baseinpore=N/2$), disagree
whether $F(N/2)$ scales logarithmically or as a small power with the
sequence length $N$.  One numerical argument in favor of a logarithmic
dependence is that different choices of the sequence disorder yield
different prefactors of the logarithm or different exponents in the
power law. While different prefactors of the logarithm are not
problematic, exponents that depend on the choice of the disorder
contradict the notion that exponents should be universal.
In~\cite{BundschuhStatisticalMechanicsRNA} the
dependence of the maximal pinch free energy on sequence length and
temperature was studied in detail and it was found that the dependence
of the maximal pinch free energy on sequence length can be described
rather well by a logarithmic law for all temperatures. At high
temperatures, the prefactor $a(T)$ of the logarithmic dependence is
$\frac{3}{2}k_B T$, as expected. However, at low temperatures this
prefactor ceases to be proportional to temperature and converges
toward a finite value at zero temperature.  Thus, if we assume that
the entire averaged pinch free energy landscape still has the
logarithmic form of (\ref{eq_logarithmiclandscape}) in the glassy
phase, the logarithmic dependence of its maximum on sequence length
implies that the effective prefactor
\begin{equation}
\label{eq_gammafroma}
\gamma(T)=\frac{a(T)}{k_B T}
\end{equation}
equals $3/2$ at high temperatures and diverges as the temperature is lowered below the glass transition temperature. For random sequences with equal probability for all four bases and energies $\ematch=\emismatch$ this behavior is numerically
illustrated in figure~\ref{fig_prefactorvstemperature}.

\section{Results and discussion}
\label{sec_results}

\subsection{Numerical analysis}
\label{sec_numerics}

From the arguments in section~\ref{sec-methods}, one may expect that the translocation of structured
RNA molecules can be described as a one-dimensional diffusion process
in the logarithmic energy landscape
(\ref{eq_logarithmiclandscape}) with a temperature dependent,
potentially large prefactor $\gamma$. Of course, the scaling of the
translocation time with sequence length in such a landscape can be
derived analytically as a function of the prefactor $\gamma$.
However, there are several uncertainties in this description. First,
different numerical studies of the pinch free energy of random RNA
molecules do not even agree on the question if the maximum of the
landscape at $\baseinpore=N/2$ scales logarithmically or with a small
power of the sequence length in the glass
phase~\cite{BundschuhStatisticalMechanicsRNA,KrzakalaMezardMuller2002,LeihanRNAGlass2006}. Second,
even if the maximum of the landscape scales logarithmically, it has
not been established that the whole average landscape follows the
simple shape (\ref{eq_logarithmiclandscape}). Third, even if the
{\em average} landscape has the suggested shape, the landscape of any
given RNA molecule can differ significantly from the average
landscape. Thus, it is not obvious how the ensemble of translocation
times of actual landscapes is related to the translocation time over
the average landscape.

To clarify these points, we perform a detailed numerical study of the
translocation dynamics of random RNA molecules on the basis of the
model defined in sections \ref{sec-translocation-1} and
\ref{sec-translocation-2}.  We generated free energy landscapes for
$2500$ different RNA molecules of length $N=1600$ using the partition
function recursion\footnote{During the calculation, the auxiliary
  partition functions $Z_{i,j}$ are rescaled to avoid numerical
  overflows due to the large exponential factors at low
  temperatures.}, (\ref{eq_rnarecursion}), and the definition
(\ref{pinchFE}) of the pinch free energy. The RNA sequences are random
and uncorrelated, with equal probabilities of $1/4$ for each of the
four bases. We use the energy model (\ref{eq_matchmismatchenergies})
with $\ematch=\emismatch$ and quote all energies in units of
$\ematch$.

\begin{figure}
\begin{indented}
\item[]\includegraphics[width=0.8\columnwidth]{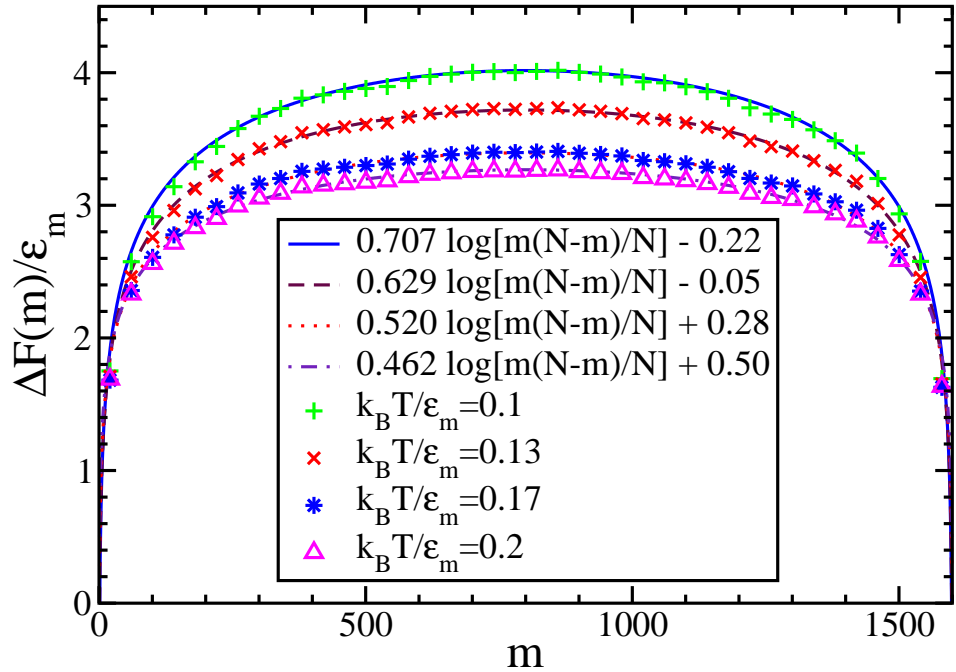}
\end{indented}
\caption{Numerically determined average free energy landscapes for
  translocation of RNA molecules through a nanopore. For clarity, the
  numerically determined free energies are averaged over ranges of the
  reaction coordinate $\baseinpore$ of size $40$. The statistical
  errors on the data are smaller than the size of the symbols.  It can
  be seen that the average free energy landscape follows the
  logarithmic shape (\protect\ref{eq_logarithmiclandscape}) with
  prefactors \protect$\gamma k_BT$ where \protect$\gamma$ is taken from
  figure~\protect\ref{fig_prefactorvstemperature} up to irrelevant
  additive constants.\label{fig_landscapes}}
\end{figure}

First of all, our ensemble of free energy landscapes allows us to
directly inspect the shape of the average
landscape. Figure~\ref{fig_landscapes} shows the pinch free energy
landscape $F(\baseinpore)$ averaged
over the $2500$ realizations of the random sequences (symbols).
Superimposed as lines are logarithmic energy landscapes as given by
(\ref{eq_logarithmiclandscape}) with prefactors $\gamma$ that are
directly obtained by multiplying the values shown in
figure~\ref{fig_prefactorvstemperature} by $k_BT$. These energy
landscapes are shifted by fitted offsets which reflect the behavior of
the pinching free energy at very small $\baseinpore$ and which are
irrelevant for the scaling behavior of the translocation dynamics. The
comparison indicates that the overall shape of the average free energy
landscapes is indeed the logarithmic one, even in the glassy
temperature regime of figure~\ref{fig_prefactorvstemperature} where
$\gamma(T)$ is significantly larger than $\frac{3}{2}$.

\begin{figure}
\begin{indented}
\item[]\includegraphics[width=0.8\columnwidth]{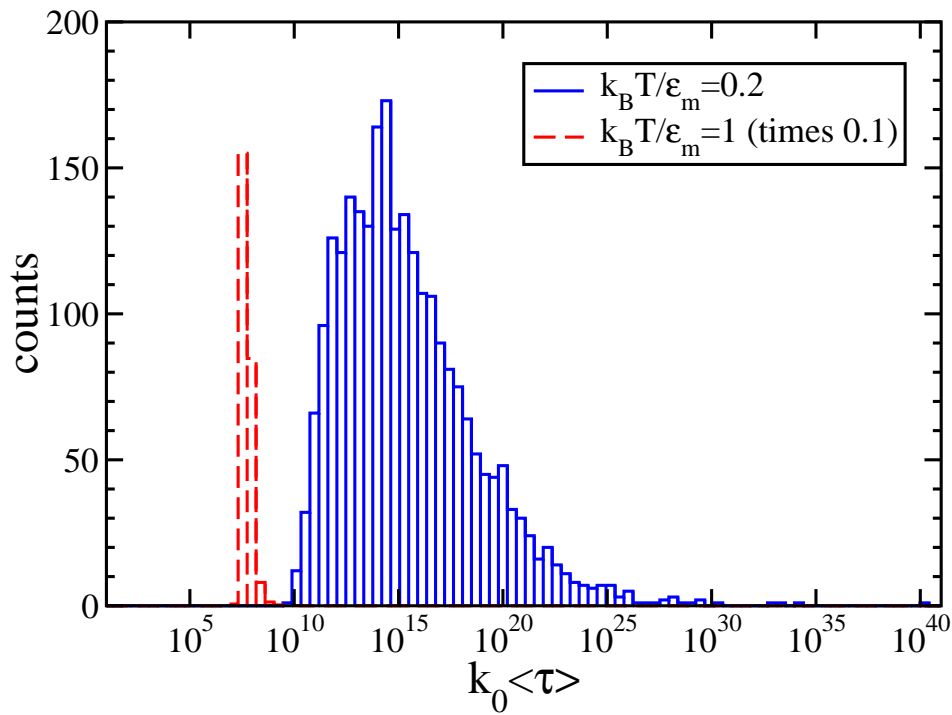}
\end{indented}
\caption{Histogram of thermally averaged translocation times for
  $2500$ random sequences of length $N=1600$ at temperatures
  $k_BT/\ematch=0.2$ and $k_BT/\ematch=1$. Note, that counts in the
  histogram for $k_BT/\ematch=1$
  are rescaled by a factor of $10$ to fit into the same plot as the
  histogram for $k_BT/\ematch=0.2$. It can
  be seen that the distribution already at $k_BT/\ematch=1$ spans
  several decades. At the low temperature $k_BT/\ematch=0.2$ the
  distribution develops a very fat tail consisting of few sequences
  with very long translocation times.\label{fig_tauhist}}
\end{figure}

Next, we examine the translocation times. For each sequence, we
calculate the thermal average of the translocation time
$\langle\tau\rangle$ using the exact expression (\ref{EQmeantau}). The
most straightforward quantity to extract from the 2500 translocation
times thus obtained would be the ensemble averaged translocation time
$\overline{\langle\tau\rangle}$, where the bar denotes averaging over
the sequence ensemble. However, as frequently observed in disordered
systems, the distribution of characteristic times develops a fat tail
at low temperatures, which renders the ensemble averaged translocation
time ill defined (see figure~\ref{fig_tauhist}). Instead, we must use a
definition for the typical value that does not rely on the existence
of the mean value. For instance, the median or the average of the
logarithm both provide a well-defined typical value even for fat
tailed distributions. Here, we use the average of the logarithm of the
translocation time, which can be interpreted as a typical effective
energy barrier.

\begin{figure}
\begin{indented}
\item[]\includegraphics[width=0.8\columnwidth]{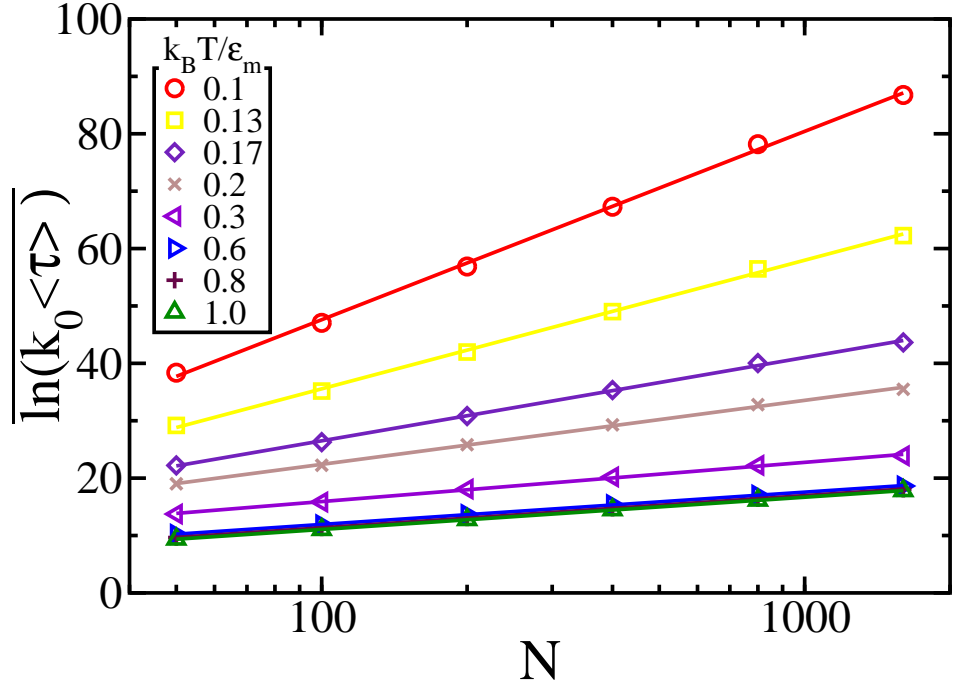}
\end{indented}
\caption{Dependence of the typical translocation time on sequence
  length for several different temperatures. To avoid problems with
  fat tails of the distribution of the translocation times at low
  temperatures, the ensemble average of the logarithms of the
  translocation times is taken to determine the typical translocation
  time. The statistical errors on the translocation times are on the
  order of the size of the symbols. It can be seen that the typical
  translocation time has a very clean power law dependence over the
  whole range of sequence lengths and for all temperatures. The
  translocation time is independent of temperature for large
  temperatures and becomes very sensitive to temperature at low
  temperatures.\label{fig_logscaling}}
\end{figure}

Figure~\ref{fig_logscaling} shows the resulting ensemble averages
of the logarithms of the translocation times for different sequence
lengths of $N=50$, $100$, $200$, $400$, $800$, and $1600$ and for
temperatures of $k_B T/\ematch=0.1$, $0.13$, $0.17$, $0.2$, $0.3$, $0.6$, $0.8$, and $1.0$. Two features of figure~\ref{fig_logscaling} are immediately obvious. First, at all temperatures, the double logarithmic plot of translocation times as a function of sequence length is perfectly linear over the whole range of sequence lengths studied. Second, the slope of these lines is independent of temperature for large temperatures and increases sharply as the temperature is lowered.

To quantify the power laws and their temperature dependence, we
apply linear regression to our logarithmic data. The resulting slopes (i.e., exponents of the power law) are shown in table~\ref{tab_exponents}. It can be seen that {\em all} exponents are larger than two, i.e. the trivial diffusive scaling $\tau\sim N^2$ does not describe the translocation dynamics. In the high temperature regime, the exponent is clearly independent of temperature, while it becomes large and very sensitive to temperature in the low temperature regime. This salient feature in the translocation behavior of structured RNA molecules implies highly anomalous sub-diffusive dynamics for the translocation process. In the next section, we will discuss this behavior from a theoretical perspective. 

\subsection{Anomalous scaling of the translocation times}
\label{sec_discussion}

Given that the translocation dynamics on a free energy landscape of
the logarithmic form~(\ref{eq_logarithmiclandscape}) was previously
studied, and its scaling behavior was found to be normally
diffusive~\cite{ChuangKantorKardarAnomalousDynamics}, our finding of
anomalous scaling in the present case is surprising. Our numerical
computation of the {\em average} free energy landscape shown in
figure~\ref{fig_landscapes} indeed confirmed that the typical landscape
for the translocation of structured RNA molecules has the logarithmic
shape, as the theoretical arguments of Sec.~\ref{sec-translocation-2}
had suggested. To resolve the apparent contradiction, we now revisit
the arguments of reference~\cite{ChuangKantorKardarAnomalousDynamics}
that led to the diffusive scaling.

\begin{table}
\caption{Exponents of the power law dependence of the typical
  translocation time on the length of the molecule. These exponents
  are determined by linear regression of the data in
  figure~\protect\ref{fig_logscaling}.\label{tab_exponents}}
\lineup
\begin{indented}
\item[]\begin{tabular}{@{}ll}
\br
$k_BT/\ematch$&exponent\\
\mr
$0.1$&$14.05\pm0.08$\\
$0.13$&\0$9.62\pm0.03$\\
$0.17$&\0$6.23\pm0.04$\\
$0.2$&\0$4.79\pm0.06$\\
$0.3$&\0$2.94\pm0.06$\\
$0.6$&\0$2.44\pm0.02$\\
$0.8$&\0$2.43\pm0.01$\\
$1.0$&\0$2.44\pm0.01$\\
\br
\end{tabular}
\end{indented}
\end{table}

Chuang, Kantor, and Kardar considered a continuum description of the
translocation process, based on a Fokker-Planck equation similar to
(\ref{Eq_Drift_Diffusion}), but with the drift velocity $v$
replaced by the local gradient of the free energy landscape,
\begin{displaymath}
\frac{\partial}{\partial t} P = D\,\frac{\partial^2}{\partial x^2} P + \frac{D}{k_B T} 
\frac{\partial}{\partial x} \left(P \,\frac{\partial}{\partial x} F(x)\right)\;, 
\end{displaymath}
with $F(x)=\gamma\, k_BT \ln [(N-x)x/N]$. They note that the polymer length $N$ and 
the diffusion constant can be eliminated from this equation by introducing a rescaled time $\tau = t D/N^2 $ and translocation coordinate $s = x/N$,
\begin{equation}
\label{dimlessFP}
\frac{\partial p}{\partial \tau} = \frac{\partial^{2}p}{\partial s^{2}} + \gamma \frac{\partial}{\partial s} \left( \frac{1-2s}{(1-s)s}\,p \right)\;, 
\end{equation}
where $p=p(s,\tau)$ now is the probability distribution in the
rescaled variables. Consequently, the authors then argue that the
solution of this dimensionless equation may be converted back to real
time by multiplying the time axis by $N^2/D$, resulting in a diffusive
scaling of the translocation time. Indeed, as the authors point out,
the argument is independent of the value of $\gamma$. Application of
the argument to the present case, with $\gamma\ge 3/2$, would suggest
that the secondary structure of the RNA is irrelevant in the slow
translocation limit considered here.

However, we will now see that this conclusion cannot be drawn. The
argument rests on the tacit assumption that the probability
distribution $p=p(s,\tau)$ develops no structure on a microscopic
scale. For instance, the continuum Fokker-Planck description breaks
down, if most of the probability is localized on one or a few points
along the translocation coordinate $m$. Indeed, such a localization
transition occurs, if $\gamma$ exceeds a threshold value of one:
Assuming a quasi-stationary solution to (\ref{dimlessFP}) which is
localized at the $s=0$ border is a self-consistent ansatz, if $p$
behaves as $p\sim s^{-\gamma}$ for small $s$. For $\gamma>1$ the
integral of this distribution diverges at the boundary, i.e. the free
energy barrier to translocation becomes strong enough for the
quasi-stationary distribution to localize at the boundary.

In the regime $\gamma>1$ where the argument for the independence of
the translocation time on $\gamma$ breaks down, the correct scaling
behavior of the translocation time can be obtained using the standard
Kramers rate theory for thermally-induced barrier
crossing~\cite{GardinerStochasticMethods,KramersProblem}. In the
present case, this approach yields
\begin{equation}
\label{eq_Kramers}
\tau\sim \frac{e^{F(N/2)/k_BT}}{\sqrt{F''(N/2)}} \sim N^{\gamma+1} \;.
\end{equation}
It is important to note that the barrier height itself according
to (\ref{eq_logarithmiclandscape}) only yields a power law
of $N^\gamma$ and that the additional power of $N$ results from the
prefactor which is often ignored in applications of Kramers rate theory.

\begin{figure}
\begin{indented}
\item[]\includegraphics[width=0.8\columnwidth]{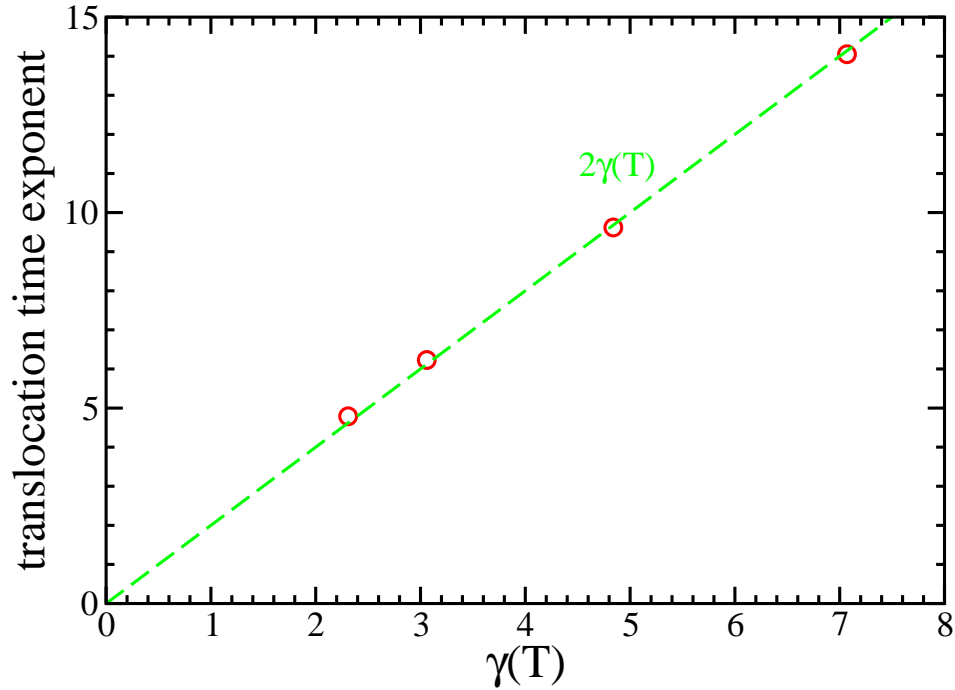}
\end{indented}
\caption{Comparison of the landscape prefactors $\gamma(T)$ from
figure~\ref{fig_prefactorvstemperature} and the numerically
determined translocation time exponents from
table~\ref{tab_exponents} for the temperatures $k_B T/\ematch$ where
RNA is expected to be in the glassy phase. It can be seen that
the observed translocation time exponents empirically behave like
$2\gamma(T)$.\label{fig_transexp}}
\end{figure}

If we apply (\ref{eq_Kramers}) to the molten phase where
$\gamma=\frac{3}{2}$, this yields $\tau\sim N^{2.5}$ in good agreement with
our numerical estimates, see table~\ref{tab_exponents}. Furthermore,
(\ref{eq_Kramers}) rationalizes the sharp increase of the scaling
exponent as the temperature is lowered into the glass phase, i.e.,
for $k_B T/\ematch\le0.2$.
However, in that regime the quantitative comparison of the translocation time exponents
in table~\ref{tab_exponents} and the barrier heights in
figure~\ref{fig_prefactorvstemperature} shown in figure~\ref{fig_transexp}
reveals that the translocation
time exponents increase even more dramatically than the increase of
the landscape prefactors suggests, namely approximately like
$2\gamma(T)$. This indicates that the typical translocation of individual
molecules in the glass phase of RNA is {\em not} well approximated by
translocation in the average landscape but rather must be dominated by
the fluctuations of the free energy landscape around the average.

\section{Conclusion and outlook}
\label{sec_conclusion}

In conclusion, we see that our numerical observation that the scaling
of the typical translocation time is drastically affected by the
secondary structure, is qualitatively well in accord with theoretical
expectation but quantitatively even exceeds the magnitude of the
effect expected from the theory. For translocation in a logarithmic
landscape it is clear from (\ref{eq_Kramers}) that
$\gamma=1$ constitutes a threshold for a change in the translocation
behavior: The regime $\gamma<1$ is marked by an insignificant barrier,
diffusive translocation, and failure of the Kramers approximation,
which assumes a ``reaction-limited'' process and ignores the time
required to diffuse from the starting to the end point. In contrast,
for $\gamma>1$ the barrier dominates the translocation dynamics and
leads to the sub-diffusive scaling (\ref{eq_Kramers}).  Importantly,
for unstructured polymers where the logarithmic free energy landscape
is only due to the configurational entropy of the polymer, we have
$\gamma<1$ even if self-avoidance is included. Here, we found that the
case of structured RNA molecules is always in the opposite regime of
$\gamma>1$. Thus, despite the similarity in the form of the free
energy landscape, (\ref{eq_logarithmiclandscape}), the
translocation behavior of unstructured and structured polynucleotides
is quite different.

The anomalous scaling of translocation times found in our study is
only observable in the absence of an external voltage. In the presence
of an external voltage the gain in electrostatic energy due to moving
$N/2$ bases into the pore is linear in $N$ and thus for large $N$
always overcomes the logarithmic barrier
(\ref{eq_logarithmiclandscape}) leading to a linear dependence of
the translocation time on sequence length. However, for finite but
small voltages the anomalous scaling could still be observed in an
intermediate regime of sequence lengths where $Nq_{\rm
  eff}V/2<\gamma(T)$.

Our empirical finding of even stronger anomalous scaling in the glassy
phase than expected from the average free energy landscape indicates
that translocation in the glassy phase is strongly affected by the
fluctuations and the free energy landscapes of the individual RNA
molecules. Understanding these fluctuations and the origin of the
intriguing empirically found $2\gamma(T)$ law for the translocation
time exponent will be subject of future research.

\ack

We gratefully acknowledge stimulating discussions with Yariv Kafri and
Julius Lucks. This work was supported in part by the Petroleum
Research Fund of the American Chemical Society through Grant
No.~42555-G9 to RB, by the National Science Foundation through grant
No.~DMR-0706002 to RB, and by the Deutsche Forschungsgemeinschaft via
SFB 486 to UG.

\section*{References}

\bibliography{anomaltrans}
\bibliographystyle{unsrt}

\end{document}